\documentclass[aps,epsf,twocolumn,showpacs]{revtex4}
\usepackage{amsmath}
\usepackage{epsfig}

\begin{document}

\title{Strong violation of critical phenomena universality:\\
Wang-Landau study of the 2d Blume-Capel model under bond
randomness}

\author{A. Malakis$^1$}

\author{A. Nihat Berker$^{2,3,4}$}

\author{I. A. Hadjiagapiou$^1$}

\author{N. G. Fytas$^1$}

\affiliation{$^1$Department of Physics, Section of Solid State
Physics, University of Athens, Panepistimiopolis, GR 15784
Zografos, Athens, Greece}

\affiliation{$^2$College of Sciences and Arts, Ko\c{c} University,
Sar\i yer 34450, Istanbul, Turkey}

\affiliation{$^3$Department of Physics, Massachusetts Institute of
Technology, Cambridge, Massachusetts 02139, U.S.A.}

\affiliation{$^4$Feza G\"{u}rsey Research Institute,
T\"{U}B\.{I}TAK - Bosphorus University, \c{C}engelk\"{o}y 34684,
Istanbul, Turkey}

\date{\today}

\begin{abstract}
We study the pure and random-bond versions of the square lattice
ferromagnetic Blume-Capel model, in both the first-order and
second-order phase transition regimes of the pure model. Phase
transition temperatures, thermal and magnetic critical exponents
are determined for lattice sizes in the range $L=20-100$ via a
sophisticated two-stage numerical strategy of entropic sampling in
dominant energy subspaces, using mainly the Wang-Landau algorithm.
The second-order phase transition, emerging under random bonds
from the second-order regime of the pure model, has the same
values of critical exponents as the 2d Ising universality class,
with the effect of the bond disorder on the specific heat being
well described by double-logarithmic corrections, our findings
thus supporting the marginal irrelevance of quenched bond
randomness. On the other hand, the second-order transition,
emerging under bond randomness from the first-order regime of the
pure model, has a distinctive universality class with
$\nu=1.30(6)$ and $\beta/\nu=0.128(5)$. These results amount to a
strong violation of universality principle of critical phenomena,
since these two second-order transitions, with different sets of
critical exponents, are between the same ferromagnetic and
paramagnetic phases. Furthermore, the latter of these two sets of
results supports an extensive but weak universality, since it has
the same magnetic critical exponent (but a different thermal
critical exponent) as a wide variety of two-dimensional systems
with and without quenched disorder. In the conversion by bond
randomness of the first-order transition of the pure system to
second order, we detect, by introducing and evaluating
connectivity spin densities, a microsegregation that also explains
the increase we find in the phase transition temperature under
bond randomness.

\end{abstract}

\pacs{75.10.Nr, 05.50.+q, 64.60.Cn, 75.10.Hk} \maketitle

\section{Introduction: Strong Violation of Universality}
\label{sec:1}

Universality, according to which the same critical exponents occur
in all second-order phase transitions between the same two phases,
erstwhile phenomenologically established, has been a leading
principle of critical phenomena~\cite{Stanley}. The explanation of
universality, in terms of diverse Hamiltonian flows to a single
fixed point, has been one of the crowning achievements of
renormalization-group theory~\cite{Wilson}. In rather specialized
models in spatial dimension $d=2$, such as the
eight-vertex~\cite{Baxter} and Ashkin-Teller~\cite{Ashkin-Teller}
models, the critical exponents nevertheless vary continuously along
a line of second-order transitions. We shall refer to these cases as
the weak violation of universality. We have established in the
current study a much stronger and more general instance of
universality violation, under the effect of quenched bond
randomness. It has been known that quenched bond randomness may or
may not modify the critical exponents of second-order phase
transitions, based on the Harris criterion~\cite{harris74,berker90}.
It was more recently established that quenched bond randomness
always affects first-order phase transitions by conversion to
second-order phase transitions, for infinitesimal randomness in
$d=2$~\cite{aizenman89,hui89} and after a threshold amount of
randomness in $d>2$~\cite{hui89}, as also inferred by general
arguments~\cite{berker93}. These predictions~\cite{aizenman89,hui89}
have been confirmed by Monte Carlo simulations~\cite{chen92,chen95}.
Moreover, renormalization-group calculations~\cite{falicov96} on
tricritical systems have revealed that not only first-order
transitions are converted to second-order transitions, but the
latter are controlled by a distinctive strong-coupling fixed point.

In the current Wang-Landau (WL) study yielding essentially exact
information on the two-dimensional (2d) Blume-Capel model under
quenched bond randomness, we find dramatically different critical
behaviors of the second-order phase transitions emerging from the
first- and second-order regimes of the pure model. These
second-order transitions with the different critical exponents are
between the same two phases indicating a strong violation of
universality, namely different sets of critical exponents on two
segments of the same critical line. Moreover, the effect of quenched
bond randomness on the critical temperature is opposite in these two
regimes, which we are able to explain in terms of a microsegregation
mechanism that we observe. Finally, in proving a general strong
violation of universality under quenched bond randomness, our study
supports a more delicate and extensive weak
universality~\cite{suzuki74,gunton75}: In the random-bond
second-order transition emerging from the pure-system first-order
transition, the magnetic (but not thermal) critical exponent appears
to be the same as that of the pure 2d Ising model, as has also been
seen in other random and non-random systems.

The Blume-Capel (BC) model~\cite{blume66,capel66} is defined by
the Hamiltonian
\begin{equation}
\label{eq:1}
H_{p}=-J\sum_{<ij>}s_{i}s_{j}+\Delta\sum_{i}s_{i}^{2},
\end{equation}
where the spin variables $s_{i}$ take on the values $-1, 0$, or
$+1$, $<ij>$ indicates summation over all nearest-neighbor pairs
of sites, and the ferromagnetic exchange interaction is taken as
$J=1$. The model given by Eq.~(\ref{eq:1}), studied here in 2d on
a square lattice, will be referred to as the pure model. Our main
focus, on the other hand, is the case with bond disorder given by
the bimodal distribution
\begin{align}
\label{eq:2}
P(J_{ij})~=~&\frac{1}{2}~[\delta(J_{ij}-J_{1})+\delta(J_{ij}-J_{2})]\;;\\
\nonumber &\frac{J_{1}+J_{2}}{2}=1\;;\;\;r=\frac{J_{2}}{J_{1}}\;,
\end{align}
so that $r$ reflects the strength of the bond randomness. The
resulting quenched disordered (random-bond) version of the
Hamiltonian defined in Eq.~(\ref{eq:1}) reads now as
\begin{equation}
\label{eq:3} H=-\sum_{<ij>}J_{ij}s_{i}s_{j}+\Delta
\sum_{i}s_{i}^{2}.
\end{equation}

\section{Two-Stage Entropic Sampling}
\label{sec:2}

We briefly describe our numerical approach used to estimate the
properties of a large number, 100, of bond disorder realizations,
for lattice sizes $L=20-100$. The pure-system properties are also
obtained, for reference and contrast. We have used a two-stage
strategy of a restricted entropic sampling, which is described in
our recent study of random-bond Ising spin models in
2d~\cite{fytas08c}, very similar to the one applied also in our
numerical approach to the 3d random-field Ising
model~\cite{fytas08a}. In these papers, we have presented in detail
the various sophisticated routes used for the restriction of the
energy subspace and the implementation of the WL
algorithm~\cite{wang01,wang01b}. The identification of the
appropriate energy subspace $(E_{1},E_{2})$ for the entropic
sampling of each random-bond realization is carried out by applying
our critical minimum energy subspace (CrMES)
restriction~\cite{malakis04,malakis05} and taking the union subspace
at both pseudocritical temperatures of the specific heat and
susceptibility. This subspace, extended by $10\%$ from both low- and
high-energy sides, is sufficient for an accurate estimation of all
finite-size anomalies. Following Ref.~\cite{fytas08c}, the
identification of the appropriate energy subspace is carried out in
the first multi-range (multi-R) WL stage in a wide energy subspace.
The WL refinement levels ($G(E)\rightarrow f * G(E)$, where $G(E)$
is the density of states (DOS); for more details see
Ref.~\cite{fytas08c}) used in this stage ($j=1,\ldots,j_{i};
f_{j+1}=\sqrt{f_{j}}$) were $j_{i}=18$ for $L< 80$ and $j_{i}=19$
for $L\geq 80$. The same process was repeated several times,
typically $\sim 5$ times, in the newly identified restricted energy
subspace. From our experience, this repeated application of the
first multi-R WL approach greatly improves accuracy and then the
resulting accurate DOS is used for a final redefinition of the
restricted subspace, in which the final entropic scheme (second
stage) is applied. In this stage, the refinement WL levels
$j=j_{i},\ldots,j_{i}+4$ are used in a one-range (one-R) or in a
multi-R fashion. For the present model, both approaches were tested
and found to be sufficiently accurate, provided that the multi-R
uses adequately large energy subintervals. This fact will be
illustrated in the following section, by presenting the rather
sensitive double-peak (dp) structure of the energy probability
density function (PDF) in the first-order regime of the model.
Noteworthy, that most of our simulations of the 2d BC model at the
second-order regime ($\Delta=1$) were carried out by using in the
final stage a one-R approach, in which the WL modification factor
was adjusted according to the rule $\ln f\sim t^{-1}$ proposed
recently by Belardinelli and Pereyra~\cite{belardinelli07}. Our
comparative tests showed that this alternative approach yields
results in agreement with the one-R WL approach.
\begin{figure*}[htbp]
\includegraphics*[width=18 cm]{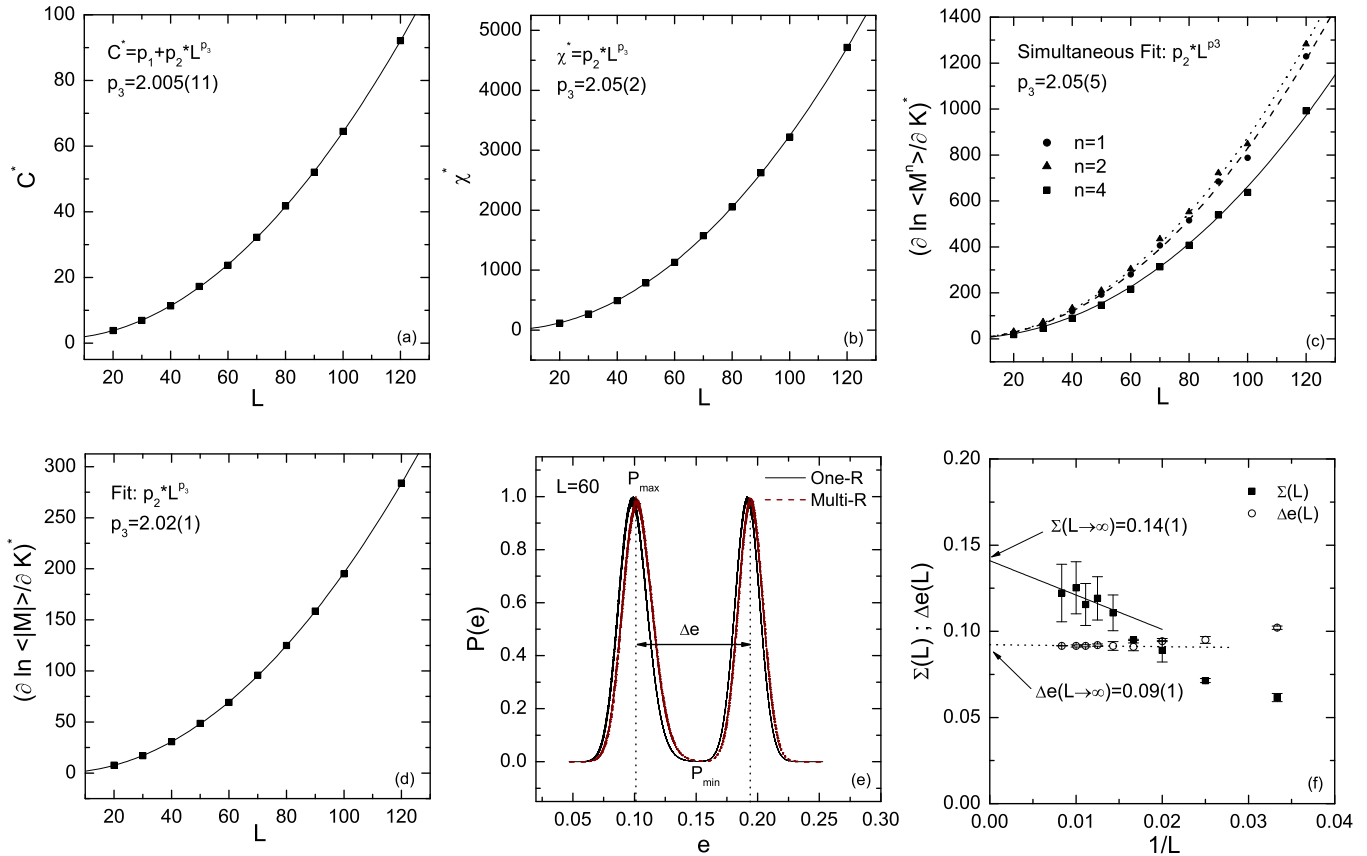}
\caption{\label{fig:1} (color online) Behavior of the pure 2d BC
model at $\Delta=1.975$: (a) FSS behavior of the specific heat
peaks giving a clear $L^{2}$ divergence. (b) The same for the
susceptibility maxima. (c) Simultaneous fitting of the maxima of
the averaged logarithmic derivatives of the order parameter
defined in Eq.~(\ref{eq:4}). (d) Power-law behavior of the
averaged absolute order-parameter derivative in Eq.~(\ref{eq:5}).
(e) The dp structure of the energy PDF at $T=T_{h}$ via the two
different implementations of the WL scheme. The multi-R
implementation is displaced very slightly to the right. (f)
Limiting behavior for the surface tension $\Sigma (L)$ defined in
the text and the latent heat $\Delta e(L)$ shown in panel (e). The
linear fits shown include only the five points corresponding to
the larger lattice sizes.}
\end{figure*}

Let us close this brief outline of our numerical scheme with some
appropriate comments concerning statistical errors and disorder
averaging. Even for the larger lattice size studied here ($L=100$),
and depending on the thermodynamic parameter, the statistical errors
of the WL method were found to be of reasonable magnitude and in
some cases to be of the order of the symbol sizes, or even smaller.
This is true for both the pure version and the individual
random-bond realizations. These WL errors have been used for the
pure system in our finite-size scaling (FSS) illustrations and
fitting attempts. For the disordered version only the averages over
the disorder realizations, denoted as $[\ldots]_{av}$, have been
used in the text and their finite-size anomalies, denoted as
$[\ldots]_{av}^{\ast}$, have been used in our FSS attempts. Due to
very large sample-to-sample fluctuations, mean values of individual
maxima ($[\ldots^{\ast}]_{av}$) have not been used in this study.
However, for the finite-size anomalies of the disordered cases, the
relevant statistical errors are due to the finite number of
simulated realizations. These errors were estimated empirically,
from runs of $20$ realizations via a jackknife method, and used in
the corresponding FSS fitting attempts. These disorder-sampling
errors may vary, depending again on the thermodynamic parameter, but
nevertheless were also found to be of the order of the symbol sizes.
For the case $\Delta=1$, these are hardly observable, as illustrated
in the corresponding graphs.

\section{Phase Transitions of the Pure $2d$ BC Model}
\label{sec:3}

\subsection{First-order transition of the pure model}
\label{sec:3a}

The value of the crystal field at the tricritical point of the
pure 2d BC model has been accurately estimated to be
$\Delta_{t}=1.965(5)$~\cite{landau81,landau86,beale86,xavier98,silva06}.
Therefore, we now consider the value $\Delta=1.975$, for which the
pure model undergoes a first-order transition between the
ferromagnetic and paramagnetic phases, and carry out a detailed
FSS analysis of the pure model.
\begin{figure*}[htbp]
\includegraphics*[width=18 cm]{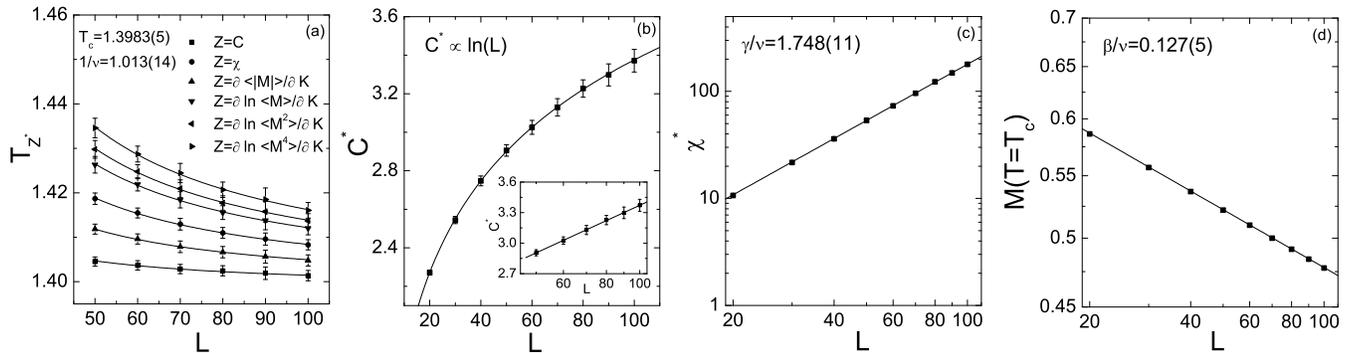}
\caption{\label{fig:2} Behavior of the pure 2d BC model at
$\Delta=1$: (a) Simultaneous fitting of six pseudocritical
temperatures defined in the text for $L\geq 50$. (b) FSS of the
specific heat peaks. The inset shows a linear fit of the specific
heat data on a log scale for $L\geq 50$. (c) FSS of the
susceptibility peaks on a log-log scale. (d) FSS of the order
parameter at the estimated critical temperature also on a log-log
scale. Linear fits are applied in panels (c) and (d).}
\end{figure*}
Our first attempt to elucidate the first-order features of the
present model will closely follow previous analogous studies
carried out on the $q=5,8,10$ Potts
model~\cite{lee90,lee91,challa86,borgs92,janke93} and also our
study of the triangular Ising model with competing nearest- and
next-nearest-neighbor antiferromagnetic
interactions~\cite{malakis07}. As it is well known from the
existing theories of first-order transitions, all finite-size
contributions enter in the scaling equations in powers of the
system size $L^{d}$~\cite{fisher82}. This holds for the general
shift behavior (for various pseudotransition temperatures) and
also for the FSS behavior of the peaks of various energy cumulants
and of the magnetic susceptibility. It is also well known that the
dp structure of the energy PDF, $P(e)$, where $e=H/L^{2}$, is
signaling the emergence of the expected two delta-peak behavior in
the thermodynamic limit, for a genuine first-order phase
transition~\cite{binder84,binder87}, and with increasing lattice
size the barrier between the two peaks should steadily increase.
According to the arguments of Lee and
Kosterlitz~\cite{lee90,lee91} the quantity $\Delta
F(L)/L=[k_{B}T\ln{(P_{max}/P_{min})}]/L$, where $P_{max}$ and
$P_{min}$ are the maximum and minimum energy PDF values at the
temperature $T_{h}$ where the two peaks are of equal height,
should tend to a non-zero value. Similarly to the above, the
logarithmic derivatives of the powers of the order parameter with
respect to the inverse temperature $K=1/T$,
\begin{equation}
\label{eq:4} \frac{\partial \ln \langle M^{n}\rangle}{\partial
K}=\frac{\langle M^{n}H\rangle}{\langle M^{n}\rangle}-\langle
H\rangle,
\end{equation}
and the average absolute order-parameter derivative,
\begin{equation}
\label{eq:5} \frac{\partial \langle |M|\rangle}{\partial
K}=\langle |M|H\rangle-\langle |M|\rangle\langle H\rangle,
\end{equation}
have maxima that scale as $L^{d}$ with the system size in the FSS
analysis of a first-order transition. In the case of a
second-order transition, the quantities in Eqs.~(\ref{eq:4}) and
(\ref{eq:5}) respectively scale as $L^{1/\nu}$ and
$L^{(1-\beta)/\nu}$~\cite{ferrenberg91,chen92,chen95}, to be used
further below.

Figures~\ref{fig:1}(a) and (b) illustrate that the traditionally
used divergences in FSS of the specific heat $C$ and susceptibility
$\chi$ follow very well a power law of the form $\sim L^{d}$, as
expected for first-order transitions~\cite{binder84,binder87}.
Furthermore, Figs.~\ref{fig:1}(c) and (d) demonstrate that the
divergences corresponding to the first-, second-, and fourth-order
logarithmic derivatives of the order parameter defined in
Eq.~(\ref{eq:4}) and the absolute order-parameter derivative defined
in Eq.~(\ref{eq:5}) follow also very well the same $L^{d}$ behavior,
as expected. Figure~\ref{fig:1}(e) shows the pronounced dp structure
of the energy PDF of the model at $T=T_{h}$ for $L=60$, obtained by
the two different implementations of the WL scheme. This graph
illustrates that the dp structure is not very sensitive to the
multi-R WL process, in contrast to our recent findings for some
first-order-like behavior of the 3d random-field Ising model. It
also illustrates the accuracy of the implementation schemes. As
mentioned above, from these dp energy PDF's one can estimate the
surface tension $\Sigma(L)=\Delta F(L)/L$ and the latent heat
$\Delta e(L)$, whose values remain finite for a genuine first-order
transition. Figure~\ref{fig:1}(f) shows the limiting behavior of
these two quantities and verifies the persistence of the first-order
character of the transition at $\Delta=1.975$. The limiting values
of $\Sigma(L)$ and $\Delta e(L)$ are given in the graph by
extrapolating at the larger lattice sizes studied. We close this
section by noting that the transition temperature
$T^{\ast}(\Delta=1.975)$ is estimated to be, in the limit
$L\rightarrow \infty$, $T^\ast=0.574(2)$. This value interpolates
and agrees with the general phase diagram points summarized in
Ref.~\cite{silva06}.
\begin{figure*}[htbp]
\includegraphics*[width=18 cm]{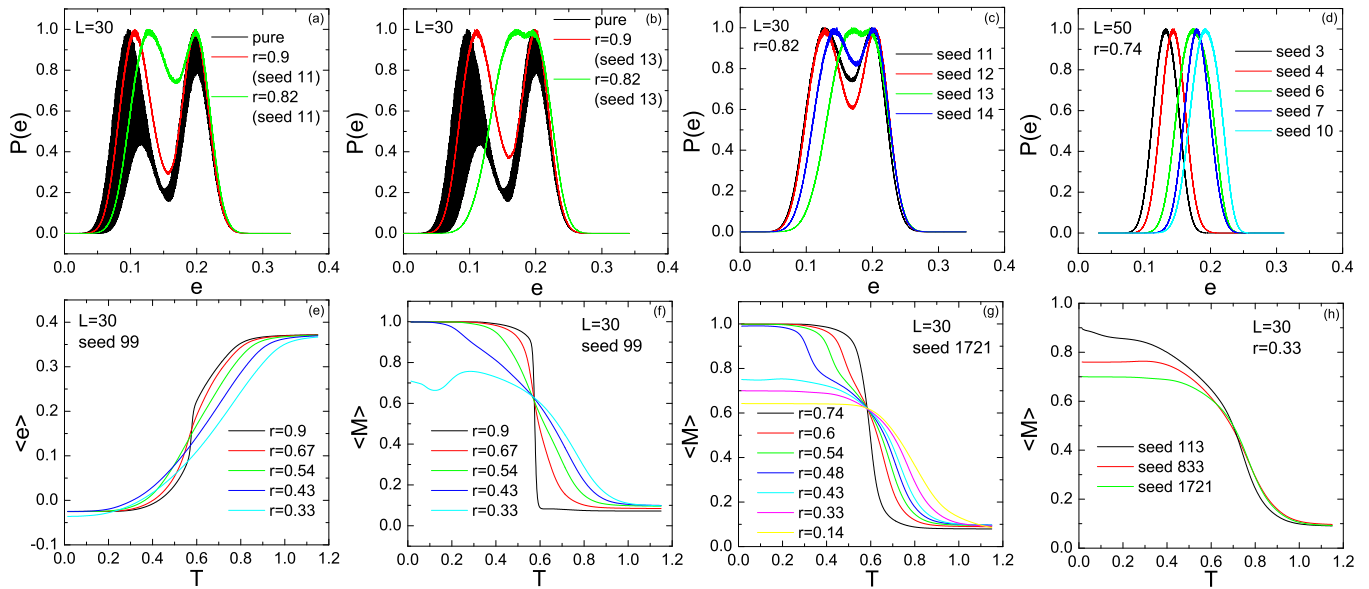}
\caption{\label{fig:3} (color online) Behavior of the random-bond
2d BC model at $\Delta=1.975$: (a)-(d): Softening effects on the
first-order transition features of the 2d BC model, induced by
bond randomness of various strengths. In (a) and (b), the very
rough $P(e)$ is for the pure model, whereas the smoothed curves
are for $r = 0.90$ (deeper) and $r = 0.82$ (shallower). The curves
in (c) and (d) are for different seeds. Panels (e) - (h)
illustrate the $\langle e \rangle$ - $T$ behavior and $\langle M
\rangle$ - $T$ behavior, for the size $L=30$ and various disorder
strengths $r$ for different disorder realizations. The curves in
(e) and (f) are for $r = 0.90,0.67,0.54,0.43,0.33$, top to bottom
on the right of (e), bottom to top on the right of (f), and top to
bottom on the left of (f). The curves in (g) are for $r =
0.74,0.60,0.54,0.48,0.43,0.33,0.14$ top to bottom on the left and
bottom to top on the right. The curves in (h) are for different
seeds.}
\end{figure*}

\subsection{Second-order transition of the pure model}
\label{sec:3b}

The 2d BC model with no quenched randomness, Eq.~(\ref{eq:1}), at
the crystal field value $\Delta=1$, undergoes a second-order
transition between the ferromagnetic and paramagnetic phases,
expected to be in the universality class of the simple 2d Ising
model~\cite{beale86}. In the following, we present the FSS analysis
of our numerical data for this case, to verify this expectation and
to set any contrast with the behavior under quenched randomness,
presented further below. Figure~\ref{fig:2}(a) gives the shift
behavior of the pseudocritical temperatures corresponding to the
peaks of the following six quantities: specific heat, magnetic
susceptibility, derivative of the absolute order parameter, and
logarithmic derivatives of the first, second, and fourth powers of
the order parameter. Fitting our data for the larger lattice sizes
($L=50-100$) to the expected power-law behavior
$T=T_{c}+bL^{-1/\nu}$, we find that the critical temperature is
$T_{c}=1.3983(5)$ and the shift exponent is $1/\nu=1.013(14)$.
Almost the same estimate for the critical temperature is obtained
when we fix the shift exponent to the value $1/\nu=1$. Thus, the
shift behavior of the pseudocritical temperatures indicates that the
pure 2d BC model with $\Delta=1$ shares the same correlation length
exponent $\nu$ with the 2d Ising model. Figure~\ref{fig:2}(b) gives
the FSS of the specific heat peaks. Here, the expected logarithmic
divergence of the specific heat is very well obtained even from the
smaller lattice sizes as shown in the main frame. The inset is a
linear fit of the specific heat data on a log scale for $L\geq 50$.
Finally, Figs.~\ref{fig:2}(c) and (d) present our estimations for
the magnetic exponent ratios $\gamma/\nu$ and $\beta/\nu$. In panel
(c) we show the FSS behavior of the susceptibility peaks on a
log-log scale. The straight line is a linear fit for $L\geq 50$
giving the estimate $\gamma/\nu = 1.748(11)$. For the estimation of
$\beta/\nu$ we use the values of the order parameter at the
estimated critical temperature ($T_{c}=1.3983$). As shown in panel
(d), on a log-log scale, the linear fit provides the estimate
$\beta/\nu=0.127(5)$. Thus, our results for the pure 2d BC model at
$\Delta=1$ are in full agreement with the findings of
Beale~\cite{beale86} and with universality arguments that place the
pure BC model in the Ising universality class.
\begin{figure*}[htbp]
\includegraphics*[width=18 cm]{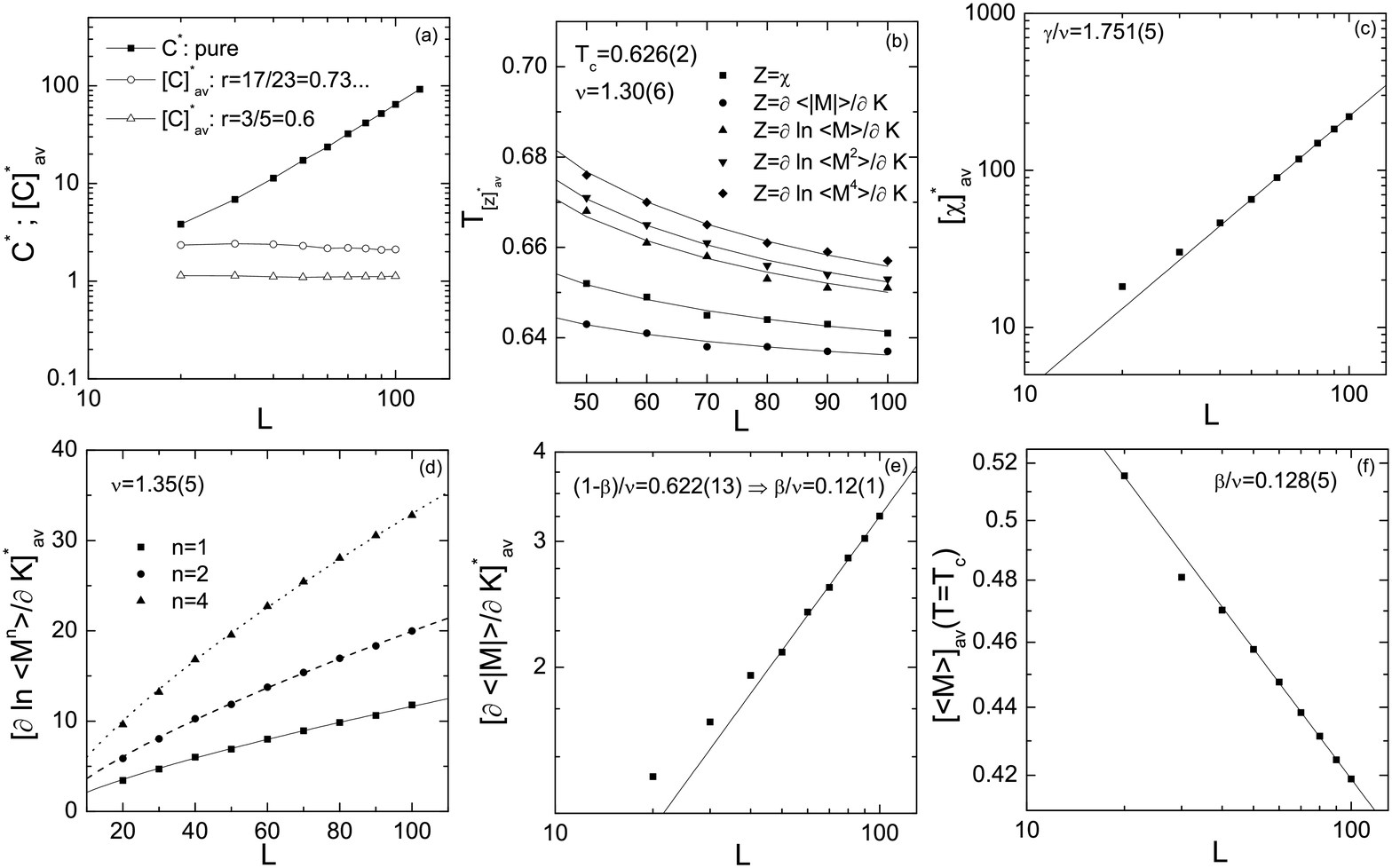}
\caption{\label{fig:4} Behavior of the random-bond 2d BC model at
$\Delta=1.975$: (a) Illustration of the clear saturation of the
specific heat ($[C]_{av}^{\ast}$) for the random-bond (open
symbols) 2d BC model. (b) FSS behavior of five pseudocritical
temperatures defined in the text for $r=3/5=0.6$. The lines show a
simultaneous fit for the larger lattice sizes $L\geq 50$. (c)-(f)
Estimation of critical exponents $\nu$, $\gamma/\nu$, and
$\beta/\nu$ for the case of $r=3/5=0.6$. In all panels the fits
have been performed for the larger lattice sizes shown ($L\geq
50$).}
\end{figure*}

\section{Phase Transitions of the Random-Bond $2d$ BC Model}
\label{sec:4}

\subsection{Second-order transition emerging under random bonds from
the first-order transition of the pure model} \label{sec:4a}

Figure~\ref{fig:3} illustrates the effects, at $\Delta=1.975$,
induced by bond randomness for different disorder realizations, on
the dp structure for lattices with linear size $L=30$
(Figs.~\ref{fig:3}(a)-(c)) and $L=50$ (Fig.~\ref{fig:3}(d)). It is
immediately seen that the introduction of bond disorder has a
dramatic influence on the dp structure of the energy PDF. The very
rough energy PDF of the pure model, with the huge oscillations
observed in relatively small lattices, is radically smoothed by the
introduction of disorder (Figs.~\ref{fig:3}(a) and (b)) and the
energy barrier is highly reduced as the disorder strength is
increased. This barrier reduction effect depends of course on the
disorder realization, as can be easily observed by comparing
Figs.~\ref{fig:3}(a), (b), and (c), but its main dependence comes
from the value of the disorder strength $r$ and already for
$r=17/23\simeq 0.74$, the dp structure is completely eliminated.
This is clarified by showing the five realizations of size $L=50$ in
Fig.~\ref{fig:3}(d). Note here that, for this value of the disorder
strength $r=17/23\simeq 0.74$, only a very small portion ($<8\%$) of
the realizations at the size $L=30$ shows a dp structure in the
energy PDF, but now with a very tiny energy barrier, whereas for the
same disorder strength, all realizations at $L=50$ have a single
peak energy PDF. We continue our illustrations of the disorder
effects by showing in Figs.~\ref{fig:3}(e) and (f) the energy
$\langle e \rangle$ - $T$ behavior and the order-parameter $\langle
M \rangle$ - $T$ behavior, for the disorder realization with seed
99. It is clear from these figures that, in effect, only the
disorder strength $r=0.9$ resembles a first-order behavior, whereas
all other disorder strengths resemble the usual second-order
behavior. In other words, only for very weak disorder strengths the
finite-size rounded anomaly resembles a discontinuity in the energy
and the order parameter.

From the order-parameter behavior shown in Fig.~\ref{fig:3}(f) it
should be observed that the low-temperature behavior, for strong
disorder strengths, shows an unexpected and rather complex behavior,
which is most prominent for $r=1/3=0.33\ldots$.
Figure~\ref{fig:3}(g) further clarifies this low-temperature effect,
by presenting one more disorder realization (seed 1721) for several
disorder strengths ranging from $r=1/7\simeq 0.14$ to $r=17/23\simeq
0.74$. It is clear from this figure that for these strong disorder
strengths the ground state of the model deviates appreciably from
the all $s_{i}=+1$ or the all $s_{i}=-1$ ferromagnetic state.
Apparently, this deviation strongly depends on the disorder strength
as shown in Fig.~\ref{fig:3}(h), where the low-temperature behavior
of the order parameter is presented for three different disorder
realizations. From this illustration it appears that, in the strong
disorder regime, the $T\rightarrow 0$ value of the order parameter
averaged over the disorder will depend on the disorder strength.
This observation will have direct relevance to the ferromagnetism
enhancement (from quenched bond disorder!) and to the conversion of
first-order transitions to second-order transitions, through the
microsegregation mechanism to be presented and quantified further
below. In fact, we have fully verified the above observation for a
small $4\times 4$ square lattice, for which the exact enumeration of
the spin configurations ($3^{16}$) is feasible, using $50$ disorder
realizations.

It is evident from the dp structures of Fig.~\ref{fig:3} that one
should avoid working with values of $J_{2}$ very close to $J_{2}=1$
(pure model), since the first-order characteristics of the pure
model may be very strong and finite-size effects will obscure any
FSS in relatively small lattices. We therefore carried out extensive
simulations at $r=17/23\simeq 0.74$ and $r=3/5=0.6$.
Figure~\ref{fig:4}(a) contrasts the specific heat results for the
pure 2d BC model and both disordered cases, $r=17/23\simeq 0.74$ and
$r=3/5=0.6$. This figure illustrates that the saturation of the
specific heat is very clear in both cases of the disorder strength.
However, the presented specific heat behavior for both disorder
strengths is unsuitable for any FSS attempt to estimate the exponent
ratio $\alpha/\nu$, as a result of the early saturation of the
specific heat. However, the early saturation of the specific heat
definitely signals the conversion of the first-order transition to a
second-order transition with a negative critical exponent $\alpha$.
Furthermore, using our numerical data, we attempted to estimate a
complete set of critical exponents for both values of the disorder
strength considered here. For $r=17/23\simeq 0.74$, our FSS attempts
indicated that we are still in a crossover regime for the lattice
sizes studied. On the other hand, for the disorder strength
$r=3/5=0.6$, the FSS attempts, using the larger lattice sizes
studied ($L=50-100$), provided an interesting and reliable set of
estimates for the critical exponents, which seems to satisfy all
expected scaling relations. Figure~\ref{fig:4}(b) gives the
behaviors of five pseudocritical temperatures $T_{[Z]^{\ast}_{av}}$
corresponding to the peaks of the following quantities averaged over
the disorder realizations: susceptibility, derivative of the
absolute order parameter, as defined in Eq.~(\ref{eq:5}), and
first-, second-, and fourth-order logarithmic derivatives of the
order parameter, as defined in Eq.~(\ref{eq:4}). The five lines
shown are obtained via a simultaneous fitting of the form
$T_{[Z]^{\ast}_{av}}=T_{c}+bL^{-1/\nu}$ for the larger lattice sizes
$L\geq 50$. The overall shift behavior is very convincing of the
accuracy of our numerical method. This accuracy is due to the fact
that for each realization, the WL random walk has been repeated in
the first stage of the entropic scheme five times, thus reducing
significantly the statistical errors, which are then further refined
in the second stage of the entropic process. Furthermore, since
these points are derived from the peaks of the averaged curves and
not from the individual maxima of the realizations, they do not
suffer from sample-to-sample fluctuations and large statistical
errors. This good behavior allows us to estimate, as shown in
Fig.~\ref{fig:4}(b), quite accurately both the critical temperature
$T_{c}=0.626(2)$ and the correlation length exponent $\nu=1.30(6)$.
Regarding the latter, we shall see below that it agrees with the
estimate obtained via the FSS of the logarithmic derivatives of
Eq.~(\ref{eq:4}) and this will be a very strong indication of the
self-consistency of our scheme.

Figures~\ref{fig:4}(c)-(f) give the FSS behavior of the first-,
second-, and fourth-order logarithmic derivatives of the order
parameter defined in Eq.~(\ref{eq:4}), the magnetic susceptibility,
the absolute order-parameter derivative defined in Eq.~(\ref{eq:5}),
and the order parameter at the critical temperature estimated in
Fig.~\ref{fig:4}(b). Figure~\ref{fig:4}(c) shows a simultaneous fit
for three moments and for lattice sizes $L\geq 50$. The resulting
value for the exponent $\nu=1.35(5)$ indeed agrees with our earlier
estimate from the shift behavior in Fig.~\ref{fig:4}(b) and also
fulfils the Chayes \emph{et al.} inequality $\nu\geq
2/d$~\cite{chayes86}. Figure~\ref{fig:4}(d) presents the behavior of
the peaks of the average susceptibility on a log-log scale and the
solid line shows a linear fit for sizes $L\geq 50$. The estimated
value for the exponent ratio $\gamma/\nu$ shown in this panel is
very close to $1.75$ and it is well known that this value of the
ratio $\gamma/\nu$ is obeyed not only in the simple Ising model but
also in several other cases in 2d. In particular, it appears that it
is very well obeyed in the cases of disordered models, including the
site-diluted, bond-diluted, and random-bond Ising
model~\cite{fytas08c,grinstein76,fisch78,dotsenko81,shalaev84,shankar87,ludwig87,kim94,dotsenko95,reis96,ballesteros97,selke98,wang90,mazzeo99,martins07,hadjiagapiou08,hasenbusch08}.
Furthermore, it has been shown that is also very well obeyed in both
the pure and random-bond version of the square Ising model with
nearest- and next-nearest-neighbor competing
interactions~\cite{fytas08c,fytas08b}, as well as the case of the
second-order phase transition induced by bond disorder from the
first-order behavior in the $q=8$ Potts model~\cite{chen92,chen95}.
Figure~\ref{fig:4}(e) is a first estimate for the exponent ratio
$\beta/\nu=0.12(1)$ obtained from the FSS behavior of the maxima of
the average absolute order-parameter derivative [Eq.~(\ref{eq:5})]
with the solid line shown being a linear fit, again for $L\geq 50$.
Finally, Fig.~\ref{fig:4}(f) shows the conventional FSS method of
estimating the ratio $\beta/\nu$ by considering the scaling behavior
of the average order parameter at the estimated critical temperature
$T_{c}=0.626$. The solid line is a linear fit for $L\geq 50$ giving
the value $\beta/\nu=0.128(5)$. These latter two estimates are very
close to the value $\beta/\nu=0.125$ and combining the above results
one finds that the random-bond version of the model appears to
satisfy the scaling relation $2\beta/\nu+\gamma/\nu=d$. Thus, a kind
of weak universality appears~\cite{kim94,kim96}.
\begin{figure*}[htbp]
\includegraphics*[width=18 cm]{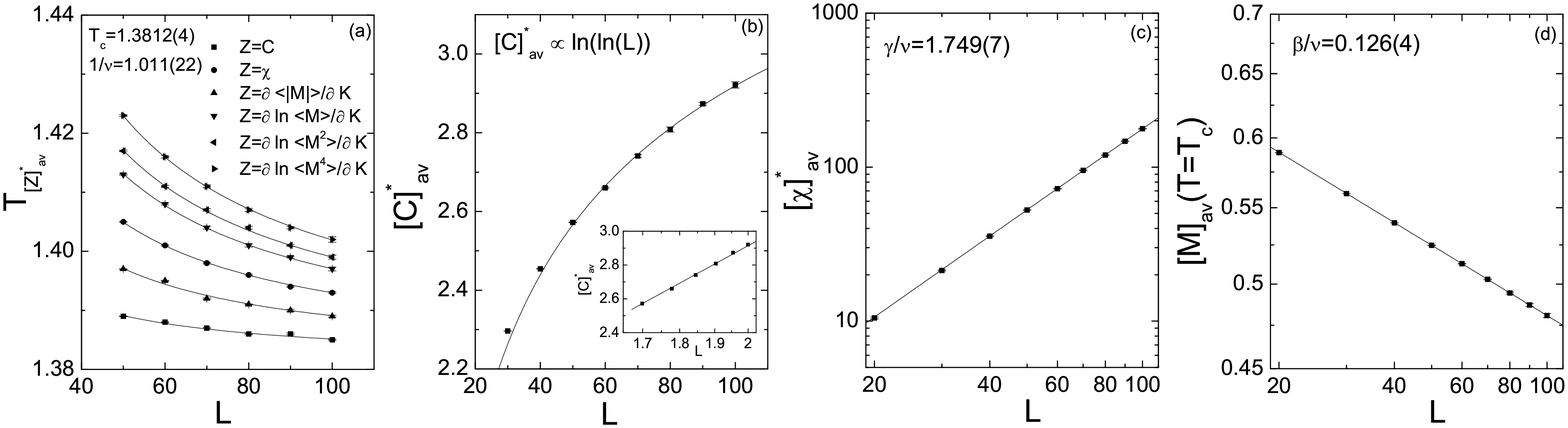}
\caption{\label{fig:5} Behavior of the random-bond ($r=0.6$) 2d BC
model at $\Delta=1$: (a) Simultaneous fitting of six pseudocritical
temperatures defined in the text for $L\geq 50$. (b) FSS of the
averaged specific heat peaks. A double-logarithmic fit is applied
for $L\geq 50$. The inset shows a linear fit on a double-logarithmic
scale. (c) FSS behavior of the averaged susceptibility peaks on a
log-log scale. (d) FSS of the averaged order parameter at the
estimated critical temperature, also on a log-log scale. Linear fits
are applied for $L\geq 50$.}
\end{figure*}

\subsection{Second-order transition emerging under random bonds from
the second-order transition of the pure model} \label{sec:4b}

We now present our numerical results for the random-bond 2d BC
model with $\Delta=1$ for disorder strength $r=0.6$. Bond
randomness favoring second-order transitions, this system is also
expected to undergo a second-order transition between the
ferromagnetic and paramagnetic phases and it is reasonable to
expect that this transition will be in the same universality class
as the random-bond 2d Ising model. As far as we know, there has
not been any previous attempt to compare the behaviors of the
random-bond BC model and the random-bond Ising model. The latter
model is a particular case of the more general random Ising model
(random-site, random-bond, and bond-diluted) and has been
extensively investigated and
debated~\cite{fytas08c,grinstein76,fisch78,dotsenko81,shalaev84,shankar87,ludwig87,kim94,dotsenko95,reis96,ballesteros97,selke98,wang90,mazzeo99,martins07,hadjiagapiou08,hasenbusch08}.
Using renormalization-group and conformal field theories, the
marginal irrelevance of randomness at the second-order
ferromagnetic-paramagnetic transition has been predicted.
According to these theories, the effect of infinitesimal disorder
gives rise only to logarithmic corrections as the critical
exponents maintain their 2d Ising values. On the other hand, there
is not full agreement in the literature for the finite disorder
regime. Two existing scenarios are mutually exclusive: The first
view predicts that the specific heat slowly diverges with a
double-logarithmic dependence, with a corresponding
correlation-length exponent
$\nu=1$~\cite{dotsenko81,shalaev84,shankar87,ludwig87}. Another
scenario predicts a negative specific heat exponent $\alpha$
leading to a saturating behavior~\cite{kim94}, with a
corresponding correlation length exponent $\nu\geq 2/d$. Let us
now present the FSS analysis of our numerical data.

Figure~\ref{fig:5}(a) presents again the shift behavior of six
pseudocritical temperatures, as defined above for the pure model,
but here using the peaks of the corresponding quantities averaged
over the disorder realizations. Fitting our data for the larger
lattice sizes ($L=50-100$) to the expected power law behavior,
$T=T_{c}+bL^{-1/\nu}$, we find that the critical temperature is
$T_{c}=1.3812(4)$ and the shift exponent is $1/\nu=1.011(22)$.
This latter estimate is a first indication that the random-bond 2d
BC at $\Delta=1$ has the same value of the correlation-length
critical exponent as the pure version and therefore as the 2d
Ising model. Figure~\ref{fig:5}(b) illustrates the FSS of the
specific heat maxima averaged over disorder, $[C]_{av}^{\ast}$.
Using these data for the larger sizes $L\geq 50$, we tried to
observe the goodness of the fits, assuming a simple logarithmic
divergence, a double-logarithmic divergence, or a simple power
law. Although there is no irrefutable way of numerically
distinguishing between the above scenarios, our fitting attempts
indicated that the double-logarithmic scenario applies better to
our data. The double-logarithmic fit is shown in the main panel
and also in the inset of Fig.~\ref{fig:5}(b). Finally,
Figs.~\ref{fig:5}(c) and (d) present our estimations for the
magnetic exponent ratios $\gamma/\nu$ and $\beta/\nu$. In panel
(c) we show the FSS behavior of the susceptibility peaks on a
log-log scale. The straight line is a linear fit for $L\geq 50$
giving the estimate 1.749(7) for $\gamma/\nu$. For the estimation
of $\beta/\nu$ we have used the values of the order parameter at
the estimated critical temperature $T_{c}=1.3812$. This
traditional method, shown in panel (d) on a log-log scale,
provides now the estimate $\beta/\nu=0.126(4)$. From the above
findings, we conclude that, at this finite disorder strength, the
random-bond 2d BC model with $\Delta=1$ belongs to the same
universality class as the random Ising model, extending the
theoretical arguments based on the marginal irrelevance of
infinitesimal randomness. Most strikingly, it is undisputable from
our numerical results that the second-order phase transitions
emerging, under random bonds, from the first-order and
second-order regimes of the pure model, have different critical
exponents although they are between the same two phases, thereby
exhibiting a strong violation of universality. We note that, since
our bond disorder occurs as the variation of the bond strengths
that all are in any case non-zero, no Griffiths
line~\cite{griffiths} divides the paramagnetic phase here.

Finally, we discuss self-averaging properties along the two segments
(ex-first-order, Sec.~IVA, and still second-order, Sec.~IVB) of the
critical line. A useful finite-size measure that characterizes the
self-averaging property of a system is the relative variance
$R_{X}=V_{X}/[X]_{av}^{2}$, where $V_{X}=[X^{2}]_{av}-[X]_{av}^{2}$,
of any singular extensive thermodynamic property $X$. A system
exhibits self-averaging when $R_{X}\rightarrow 0$ as $L\rightarrow
\infty$, or lack of self-averaging (with broad probability
distributions) when $R_{X}\rightarrow const\neq 0$ as $L\rightarrow
\infty$. The FSS scenario of Aharony and Harris~\cite{aharony96}
describes self-averaging properties for disordered systems and has
been validated by Wiseman and
Domany~\cite{wiseman95,wiseman98a,wiseman98b} in their study of
random Ising and Ashkin-Teller models. From these papers, the
disordered system resulting in Sec.~IVB from the introduction of
bond randomness to the marginal case of the second-order transition
of the pure 2d BC model is expected to exhibit lack of
self-averaging. This expectation also agrees with our recent study
of the self-averaging properties of the 2d random-bond Ising
model~\cite{fytas08d}. In the current work, the FSS behaviors of the
relative variances, obtained from the distributions of the magnetic
susceptibility maxima ($X=\chi^{\ast}$), were observed. Their
behavior clearly indicates that the disordered systems exhibit lack
of self-averaging along both segments of the critical line, since
these relative variances show a monotonic behavior and are still
increasing at the maximum lattice sizes studied. For $\Delta=1$,
$R_{\chi^{\ast}}(L=90)\simeq 0.0011$ and
$R_{\chi^{\ast}}(L=100)\simeq 0.0014$, whereas for $\Delta=1.975$,
$R_{\chi^{\ast}}(L=90)\simeq 0.015$ and
$R_{\chi^{\ast}}(L=100)\simeq 0.016$. Thus, the latter case,
\textit{i.e.}, the ex-first-order segment, gives a much larger
effect, by a factor of $\sim 12$. A similarly stronger lack of
self-averaging was observed for the disordered system resulting from
the case of competing interactions on the square lattice, than for
the disordered system resulting from the marginal case of the simple
Ising model, in our recent study~\cite{fytas08d}. Moreover, the
discussion in Sec.~VIIA in the paper by Fisher~\cite{fisher95} is
relevant here, explaining the expectation for extremely broad
distributions near ex-first-order transitions in systems with
quenched randomness. This discussion points out also that
ex-first-order transitions may have several $\nu$
exponents~\cite{chayes86,fisher95,privman83,huse} and provides the
background for understanding why our finite-size correlation-length
exponent obeys the Chayes \emph{et al.} inequality~\cite{chayes86}.

\subsection{Contrasting random-bond behavior of critical temperatures,
connectivity spin densities, and microsegregation} \label{sec:4c}

In most spin models, the introduction of bond randomness is
expected to decrease the phase-transition temperature and in
several cases the critical temperature goes to zero at the
percolation limit of randomness ($r=0$ and $J_{2}=0$). For less
randomness, only a slight decrease is expected, if the average
bond strength is maintained. Indeed, in the second-order regime of
the pure 2d BC model, $\Delta=1$, the introduction of bond
randomness has slightly decreased the critical temperature, by
$1\%$ (Sec.~\ref{sec:4b}). On the other hand, in sharp contrast,
for the same disorder strength $r=0.6$ applied to the first-order
regime of the pure model (Sec.~\ref{sec:4a}), at $\Delta=1.975$,
we find a considerable increase of the critical temperature, by
$9\%$.

\begin{figure}[htbp]
\includegraphics*[width=9 cm]{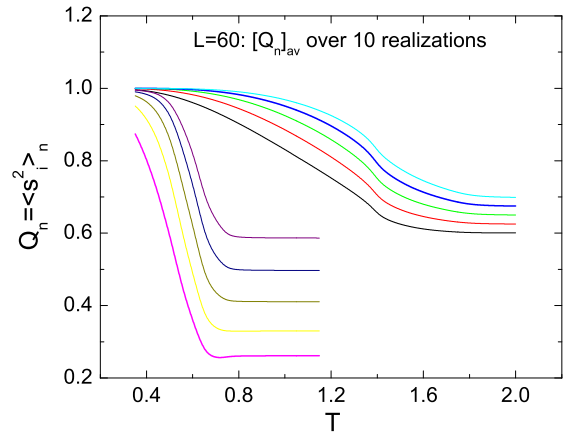}
\caption{\label{fig:6} (color online) Temperature behavior of the
connectivity spin densities $Q_{n}$ defined in the text for
$\Delta=1$ (upper curves) and $\Delta=1.975$ (lower curves) for a
lattice size $L=60$. In each group, the curves are for $n=0,1,2,3,4$
from bottom to top.}
\end{figure}

In order to microscopically explain the above observation of
ferromagnetic order enhanced by quenched disorder, let us define the
following connectivity spin densities, $Q_{n}=\langle
s_{i}^{2}\rangle_{n}$, where the subscript $n$ denotes the class of
lattice sites and is the number of the quenched strong couplings
($J_{1}$) connecting to each site in this class. Figure~\ref{fig:6}
illustrates the temperature behavior of these densities averaged
over 10 disorder realizations for a lattice of linear size $L=60$.
For $\Delta=1.975$, it is seen that the $s_{i}=0$ preferentially
occur on the low strong-coupling connectivity sites. The $s_{i}=\pm
1$ states preferentially occur with strong-coupling connectivity,
which (1) naturally leads to a higher transition temperature and (2)
effectively carries the ordering system to higher non-zero spin
densities, the domain of second-order phase transitions.
Figure~\ref{fig:6} constitutes a microsegregation, due to quenched
bond randomness, of the $s_{i}=\pm 1$ states and of the $s_{i}=0$
state. We note that microsegregation is reached by a continuous
evolution within the ferromagnetic and paramagnetic phases. A
similar mechanism has been seen in the low-temperature second-order
transition between different ordered phases under quenched
randomness~\cite{kaplan08}.

On the other hand, for $\Delta=1$ and in the neighborhood of the
critical temperature, the difference between the smallest and the
largest of the connectivity densities is $0.177$, whereas for
$\Delta=1.975$, this difference is $0.449$ in the corresponding
critical region. Thus, the microsegregation does not occur in the
regime of the second-order transition of the pure model and the
effect of quenched disorder is the expected slight (the average
bond strength is maintained) retrenchment of ferromagnetic order.
Microsegregation does occur in the first-order regime of the pure
model where macrosegregation occurs in the absence of bond
randomness. The result is a local concentration of $s_{i}=\pm 1$
states, leading to the enhancement of ferromagnetism. The
above-mentioned spreads in the connectivity-density values are
very slowly changing with the lattice size. For instance, for
$L=20$ the corresponding values are respectively $0.175$ and
$0.445$.

\section{Conclusions: Strong Violation of Universality and Weak
Universality} \label{sec:5}

In conclusion, the second-order phase transition of the 2d
random-bond Blume-Capel model at $\Delta=1$ appears to belong to
the same universality class as the 2d Ising model, having the same
values of the critical exponents, \textit{i.e.}, the critical
exponents of the 2d Ising universality class. The effect of the
bond disorder on the specific heat is well described by the double
logarithmic scenario and our findings support the marginal
irrelevance of quenched bond randomness. On the other hand, at
$\Delta=1.975$, the first-order transition of the pure model is
converted to second order, but in a distinctive universality class
with $\nu=1.30(6)$ and $\beta/\nu=0.128(5)$.

These results, on the 2d square lattice, amount to a strong
violation of universality, since the two second-order transitions
mentioned in the previous paragraph, with different sets of
critical exponents, are between the same ferromagnetic and
paramagnetic phases. This result was also obtained by
renormalization-group calculations~\cite{falicov96} in 2d and 3d
that are exact on hierarchical lattices and approximate on square
and cubic lattices. The mechanism in these renormalization-group
calculations is that the second-order transitions, emerging under
random bonds from the first-order transitions of the pure model,
have their own distinctive unstable zero-temperature fixed
point~\cite{falicov96,ozcelik}.

Furthermore, the latter of these two sets of results supports an
extensive but weak universality, since the latter of the two
transitions mentioned above has the same magnetic critical exponent
(but a different thermal critical exponent) as a wide variety of 2d
systems without~\cite{suzuki74,gunton75} and
with~\cite{fytas08c,kim94,fytas08b,kim96} quenched disorder.

\begin{acknowledgments}
This research was supported by the special Account for Research
Grants of the University of Athens under Grant No. 70/4/4071. N.G.
Fytas acknowledges financial support by the Alexander S. Onassis
Public Benefit Foundation. A.N. Berker acknowledges support by the
Academy of Sciences of Turkey.
\end{acknowledgments}

{}

\end{document}